\documentclass[aps,pra,twocolumn,showpacs,preprintnumbers,amsmath,amssymb,floatfix,superscriptaddress]{revtex4-1}

\usepackage{graphics}
\usepackage{graphicx}
\usepackage{amsmath}
\usepackage{setspace}
\usepackage{braket}
\usepackage{epstopdf}
\usepackage{comment}
\usepackage{hyperref}
\usepackage{bm}
\usepackage{xfrac}
\usepackage{xcolor}

\usepackage{mathrsfs}

\hypersetup{%
   pdfpagemode=None, 
   pdfstartpage=1,
   pdfmenubar=true,
   pdftoolbar=true,
   colorlinks = true,
   linkcolor=blue,
   citecolor=blue,
   urlcolor=blue,
   bookmarksopen=false
 }

\newcommand{\hh}  [1] {{\color{red}{#1}}}

\begin{document}
\author{Henrik Hirzler}
\address{Van der Waals-Zeeman Institute, Institute of Physics, University of Amsterdam, 1098 XH Amsterdam, The Netherlands}
\author{Jes\'us P\'{e}rez-R\'{i}os}
\address{Fritz-Haber-Institut der Max-Planck-Gesellschaft, Faradayweg 4-6, D-14195 Berlin, Germany}
\address{Institute for Molecules and Materials, Radboud University, Heyendaalseweg 135, 6525 AJ Nijmegen, The Netherlands}

\title{Rydberg atom-ion collisions in cold environments}

\date{\today}

\begin{abstract}
Low-energy collisions of Rydberg atom-ion systems are investigated theoretically. We present the parameter space associated with suitable approaches for the dynamics of Rydberg atom-ion collisions, i.e. quantum-, Langevin and classical exchange regimes, showing that for the lowest reachable temperatures a classical treatment is appropriate. A quasi-classical trajectory method is used to study charge exchange cross sections for Li$^*$-Li$^+$ and Li$^*$-Cs$^+$ at collision energies down to $1\,$K. For cold collisions we find the charge exchange cross section deviating from the $n^4$ geometric scaling. Furthermore, for low-energy collisions, we find both an influence of the ionic core-repulsion as well as variations for two different models used for describing the electron-core potential.
\end{abstract}

\maketitle

\section{Introduction}

More than a century ago, after systematically studying atomic spectral lines, Rydberg gave a glimpse into the intriguing physics of highly excited states of atoms, nowadays known as Rydberg states~\cite{Rydberg}.
Occurring in interstellar space and plasma, Rydberg states of atoms (Rydberg atoms) and their peculiar properties find significance in a diversity of subjects reaching from astrophysics to quantum computing~\cite{Gnedin2009,Vitrant1982,Wilk,Isenhower,Qsim,QCDsim,Eiles2018}. Their exaggerated properties facilitate studies of many-body physics in cold atomic samples as well as on the single atom level in optical tweezers~\cite{QuantumCritical,Schauss}. Even more exotic structures, such as ultralong-range Rydberg molecules showing a unique molecular bond caused by the elastic scattering of the Rydberg electron with a neutral atom~\cite{Pfau,Trilobite,Butterfly,Greene2000,Hamilton2002,Fabikant} and Rydberg states of charged particles have been achieved~\cite{Feldker:2015}.

New hybrid atom-ion experiments open up a path to investigate Rydberg-ion collisions at low temperatures. Two distinct approaches have crystallized: either, a \emph{quasi}-free ion is created from an ultracold atomic gas via photoionization of Rydberg atoms and subsequently measured via time-of-flight~\cite{Kleinbach_2018,Engel}, or the ion is confined inside the radio-frequency electric fields of a Paul trap~\cite{Ewald:2019,haze2019stark}. These scenarios are suitable for studying the most intriguing properties of Rydberg-ion interaction and controlling and tailoring such interaction. For instance, it is possible to engineer repulsive Rydberg-ion interaction to cool trapped ions or for realizing atom-ion quantum logic gates \cite{Secker:2016, Secker:2017}. In addition, although in a different range of collision energies, the study of charged-neutral interactions is relevant for the formation of anti-hydrogen from the scattering of positronium atoms with anti-protons~\cite{AH1,AH2,AH3}.

In fact, already in the 70's Rydberg-ion collisions were the object of an intense research, although in very different conditions. Cross-beam experiments allowed to study charge transfer between Rydberg atoms and single charged ions \cite{MacAdam_1982} as well as principal- ($n$) and angular quantum number ($l$) changing collisions~\cite{MacAdam_1985}. Similarly, combining Rydberg atoms with highly charged ion beams, e.g. Ar$^{9+}$ the collision cross section~\cite{Martin1989} and the charge exchange rate was determined~\cite{Lundeen_2001}. Furthermore, the orientation of Rydberg atoms towards the ion projectile, controlled with electric and magnetic fields, has been studied~\cite{Kohring_1983,Hare_1988}. In these pioneering works, the projectile velocity was of the same order as the average velocity of the Rydberg electron ($\sim n^{-1}$~a.u.). Thus, and due to experimental possibilities at that time, typical collision energies were limited to $\gtrsim100$\,eV~$\approx$ $ 10^{6}$~K. At the same time, a classical trajectory approach assuming a pure Coulombic interaction between the colliding particles was developed to characterize the charge-transfer and impact-ionization cross sections. However, it was primarily applied to  Ion-H$^*$ collisions~\cite{Olson}. Later, models were extended with tailored pseudo-potentials accounting for the role of core electrons in the Rydberg atom and ion, which allowed to calculate the state changing collision cross section after binning the classical phase-space at the final propagation time~\cite{Pascale}. However, none of these theoretical approaches explored the nature of the charge transfer process between an ion and a Rydberg atom at cryogenic and low temperatures.

In this paper, we present a quasi-classical trajectory (QCT) approach for charge exchange and $l$-mixing in Rydberg-ion collisions at low temperatures. The method is applicable down to temperatures $\sim 1$~K, thus exploring a novel dynamical regime in ion-Rydberg systems. Moreover, from calculating the number of contributing partial waves, we draw the \emph{parameter-space} associated with the applicability of different classical and quantum mechanical methods as a function of the collision energy and Rydberg principal quantum number. As a result, we find that for typical temperatures for Bose-Einstein condensation ($\sim 100$~nK), Rydberg-ion collisions are satisfactory described from a classical framework.

\section{A phase-diagram for scattering methods on ion-Rydberg collisions}

The long-range potential for atom-ion collisions is given by the induced dipole-charge interaction plus the centrifugal barrier as

\begin{equation}
\label{eq_init}
	V(r) = - \frac{C_4}{r^4} + \frac{L(L+1)}{2\mu r^2},
\end{equation}

\noindent
with partial wave $L$, reduced mass $\mu$ and $C_4 = \alpha(n)/2$, where 

\begin{equation}
\alpha(n_\hh{l})=a_1 (n-\delta_l)^7\left(1+\frac{a_2}{n-\delta_l}+\frac{a_3}{(n-\delta_l)^2}\right)
\end{equation}

\noindent
stands for the polarizability of the Rydberg atom in the principal quantum number $n$ and angular momenta state $l$, $a_1$, $a_2$ and $a_3$ represent asymptotic coefficients for a given atom~\cite{Kamenski:2014} and $\delta_l$ is the quantum defect for the $l$-state. 

The potential in Eq.~(\ref{eq_init}) presents a maximum that when equated to the collision energy leads to an estimation of the number of partial waves relevant for the collision under consideration as

\begin{equation}
	L \approx  2 \sqrt{\mu} C_4^{1/4} E_\mathrm{col}^{1/4}.
	\label{pw}
\end{equation}


Let us assume that for $L\gtrsim 20$ most of the relevant quantum effects on the scattering observables are washed out. Therefore, any scattering event requiring more than 20 partial waves may be treated classically. Indeed, from Eq.~(\ref{pw}) it is possible to obtain the threshold collision energy in which such condition is satisfied as 

\begin{equation}
    \label{Ecol}
    E_\mathrm{col} = \left(\frac{L}{2\sqrt{\mu}}\right)^4 C_4^{-1}.
\end{equation}

\begin{figure}[h!]
    \centering
    \includegraphics[width=\linewidth]{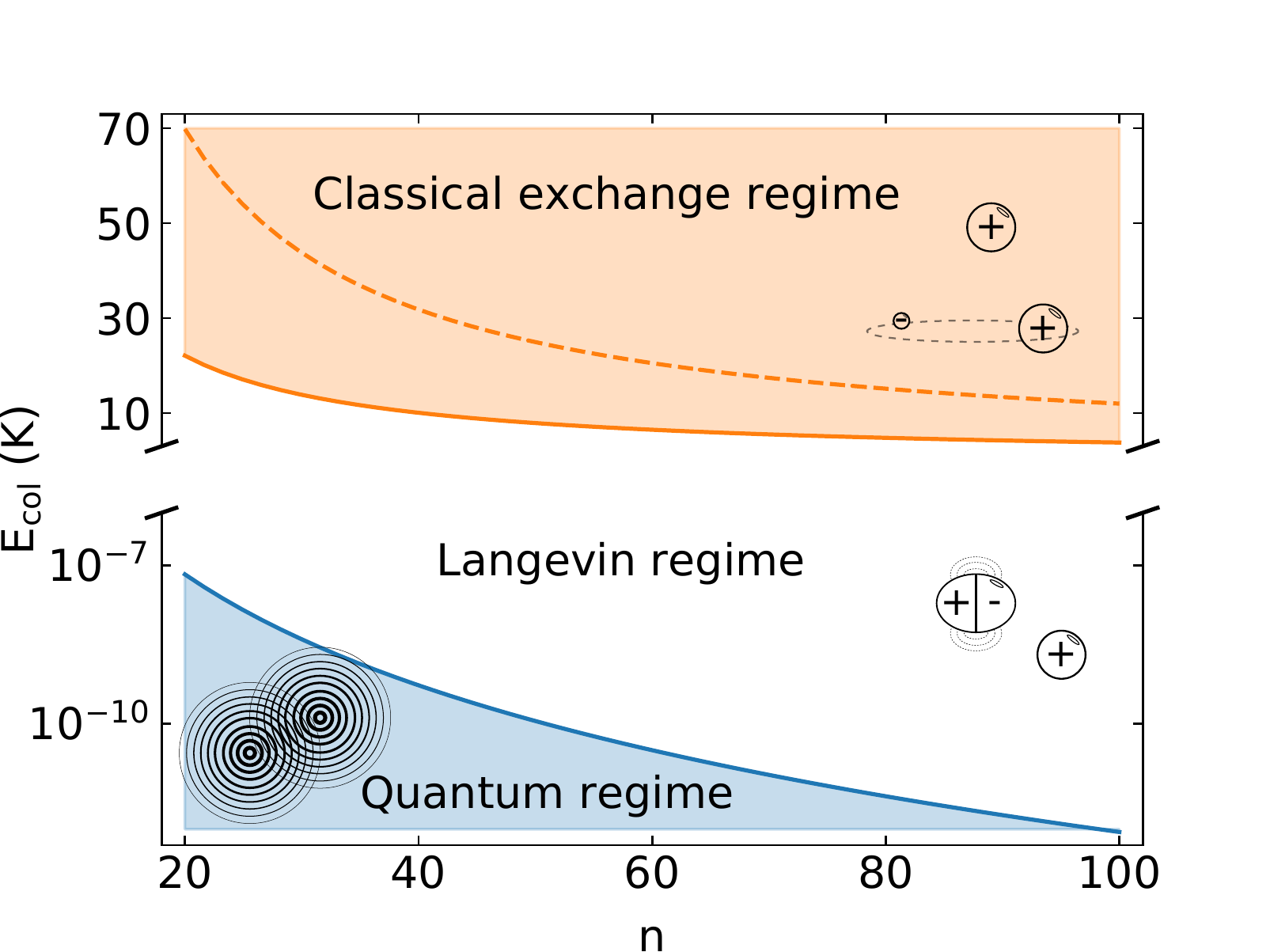}
    \caption{Parameter space associated to suitable approaches for the dynamics of Rydberg-ion collisions. The upper part of the diagram (orange in the online version) it is shown the region where a classical treatment based on Coulomb interaction is appropriate in-/excluding core-interaction (dashed/sold line). In white is shown the region where the classical treatment is adequate, but in this case assuming the expected induced dipole-charged interaction through the celebrated Langevin capture model. The region below the white one (blue in the online version) represents the parameter space in which a quantum treatment is required. The results are for Li$(nS)$ + Li$^+$ collision regarding the applicability of different approaches for the dynamics. Note, the crossover from Classical exchange to Langevin is independent of the ion species.}
    \label{phase_diagram}
\end{figure}

\noindent
Therefore, Eq.~(\ref{Ecol}) determines the borderline between the quantal and classical regimes, which is shown as the solid line in Fig.~\ref{phase_diagram}.

 \subsection{Coulomb vs Langevin}\label{CL}

\begin{figure}[h!]
    \centering
    \includegraphics[width=0.8\linewidth]{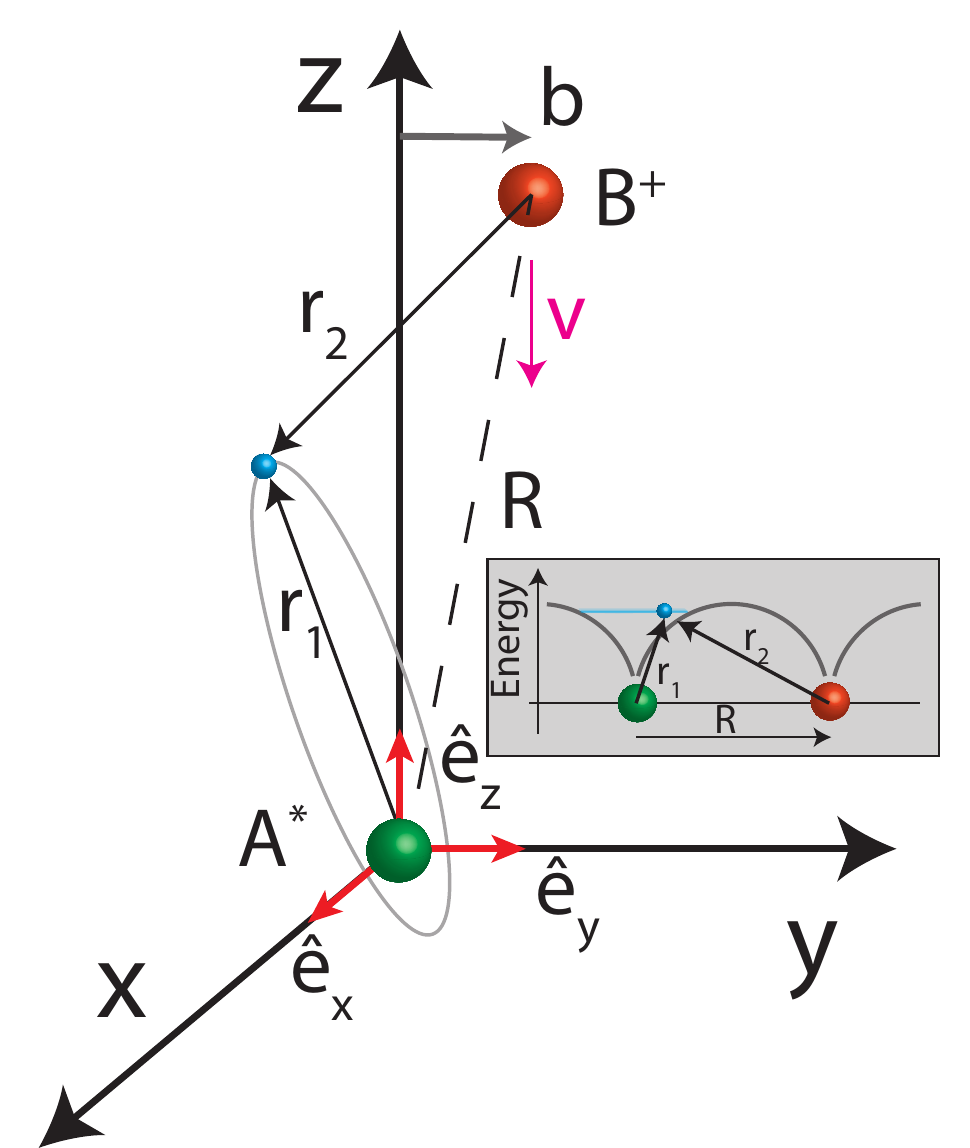}
        \caption{Schematic representation of Rydberg-ion collisions. The ion is considered the projectile with impact parameter $b$ moving with a velocity $v$ towards the Rydberg atom, placed at the origin of coordinates. The Rydberg electron's position is labeled as  ${\mathbf r}_1$ from the Rydberg core, whereas ${\mathbf r}_2$ represents its position from the ion. The inset represents a one-dimensional perspective of the charge exchange process: the electron moves through the potential barrier at a given ion-Rydberg distance. }
        \label{Fig2}
\end{figure}

The collision of a Rydberg atom and an ion may lead to a charge transfer process in which the Rydberg electron is captured by the ion, A$^*$+B$^+\rightarrow$ A$^+$+B$^*$. Let us assume, for the sake of simplicity, a resonant-charge exchange process, in which the total potential energy of the Rydberg electron (in a.u.) reads as

\begin{equation}
\label{eq1}
U({\bf r}_1,{\bf r}_2,R)=-\frac{1}{|{\bf r}_1|}-\frac{1}{|{\bf r}_2|}+\frac{1}{R},
\end{equation}

\noindent
where $R$ is the interatomic distance, ${\bf r}_1$ is the vector position of the Rydberg electron with respect to its ionic core and ${\bf r}_2$ represents the vector position of the Rydberg electron regarding the ion, as it is schematically shown in Fig.~\ref{Fig2}. When the potential energy is equal to the binding energy of the Rydberg electron $-\frac{1}{2n^{2}}$, then the charge transfer occurs. In particular, it occurs at the midpoint of the vector joining the Rydberg core and the ion, as shown in the inset of Fig.~\ref{Fig2}, and hence one finds

\begin{equation}
\label{eq2}
U(R^*/2,R^*/2,R^*)=-\frac{1}{2n^2} \rightarrow R^*=6n^2.
\end{equation}

\noindent
$R^*$ is the length scale associated with the classical charge-transfer process. Therefore, the charge exchange cross section is $\propto n^4$, which corresponds to the scaling law of the geometric cross section based on the Rydberg orbit size. It is worth noticing that the derivation is based on a barrier-less model, i.e., the collision is assumed to occur in a $s$-wave regime or at zero impact parameter. Similarly, neglecting the ion-ion repulsion term one finds that $R^*=8n^2$.

On the other hand, using the Langevin model~\cite{Langevin1905}, a Rydberg-ion scattering process, as a neutral-charge interaction system, is initially dominated by the long-range induced dipole-charge interaction, which reads as

\begin{equation}
\label{eq3}
V(r)=-\frac{\alpha(n)}{2r^4},
\end{equation}

\noindent
where we have made explicit the dependence of the Rydberg polarizability with the principal quantum number. In this scenario, the Langevin capture model is applicable, and the capture radius, know as the Langevin impact parameter, is given by

\begin{equation}
\label{eq4}
b_{\text{L}}=\left(\frac{2\alpha(n)}{E_{\text{col}}} \right)^{1/4},    
\end{equation}

\noindent
where $E_{\text{col}}$ stands for the collision energy.

The classical exchange process is energy independent, whereas the Rydberg-ion Langevin collision depends on the energy. Therefore, it is possible to find a collision energy at which the length scale matches with the one for Langevin collision

\begin{equation}
\label{eq5}
\left(\frac{2\alpha(n)}{E_{\text{col}}} \right)^{1/4}=6n^2.
\end{equation}

\noindent
This energy defines the lower bound of the validity of the classical exchange process, which is depicted as the dashed line in Fig.~\ref{phase_diagram} for Li$^*$ + Li$^+$ collisions. In addition, neglecting the ion-ion repulsion the term pushes further down the lower bound for the validity of the classical exchange process as it is shown in Fig.~\ref{phase_diagram} as the upper solid line. As a result, Li$^*$ + Li$^+$ collisions may be treated with the classical exchange process down to $E_{\text{col}}\sim 10$~K for $n\gtrsim 50$ whereas it is only valid for $E_{\text{col}}\gtrsim 30$~K for lower principal quantum numbers.

\section{Theoretical approach}\label{Sec_theory}
We use a quasi-classical trajectory (QCT) method to simulate Rydberg-ion collisions. In this approach, the dynamics of the nuclei in the electronic potential energy surface is described classically while the initial conditions for the trajectories correlate with the quantum state of the system via the semi-classical approximation. Indeed, the same procedure is employed to calculate the final quantum state of the system from the position of the trajectory in the classical phase-space~\cite{Truhlarbook,Levine,JPRBook}. The QCT method finds application in systems described by large numbers of partial waves \cite{JPR2019,JPRBook}. 

The interaction between the electron and each of the two ionic cores is described by the following pseudo potential
 
\begin{equation}
    V_\mathrm{ie}(r) = - \frac{Z(r)}{r}-\frac{\alpha_c}{2r^4}\left(1-e^{-\left(r/r_c\right)^6}  \right),
    \label{eq:param_model}
\end{equation}

\noindent
which is obtained from Ref.~\cite{Marinescu}. The first term represents a screened Coulomb interaction whereas the second stands for the interaction between the Rydberg electron and the core electrons. In particular, $\alpha_c$ is the polarizability of the core, $r_c$ stands for the radius of the core and

\begin{equation}
    Z(r) = 1+(z-1)e^{-a_1 r}-r(a_3+a_4 r)e^{-a_2 r},
\end{equation}

\noindent
where $z$ is the nuclear charge of the Rydberg atom. $a_i$ and $r_c$ are parameters taken form Ref.~\cite{Marinescu}. We note, that $a_i$ and $r_c$ have $l$ dependence. Nonetheless, we keep the parameters constant during a collision.

For the ion-ion interaction we use a coulomb potential
\begin{equation}
    V_\mathrm{ii}(R) = \mathcal{A}\frac{1}{R},
\end{equation}
where the attenuation factor $0\leq\mathcal{A}\leq 1$ controls the repulsion strength between the cores~\cite{Ostrovsky_1995}. 

Each collision is initialized by placing the Rydberg core at the origin of coordinates and the Rydberg electron in the turning point of its orbit
\begin{equation}
    r_\mathrm{1}= n^2+n\sqrt{n^2-l\left(l+1\right)},
\end{equation}
where the velocity is given by the angular momentum.

\begin{equation}
    v_\mathrm{1}= \frac{\sqrt{l\left(l+1\right)}}{n^2+n\sqrt{n^2-l\left(l+1\right)}}.
\end{equation}
We dice the direction of the position $\vec{r}_\mathrm{1}/{|\vec{r}_\mathrm{1}|}$  and the velocity $\vec{v}_\mathrm{1}/{|\vec{v}_\mathrm{1}|}$, while ensuring $\vec{r}_\mathrm{1} \cdot\vec{v}_\mathrm{1}=0$. Furthermore, the phase of the electron orbit is randomized.

The ion is initialized at $\vec{R}_0 = b \hat{e}_y + R_{0,z} \hat{e}_z$, where $b$ is the impact parameter, with velocity $-v \hat{e}_z$, as visualized in Fig.~\ref{Fig2}. $R_{0,z}$ is chosen sufficiently large, such that the Coulomb potential is less than $10\%$ of the collision energy $E_\mathrm{col} = \frac{1}{2}\mu v^2$ and in addition $R_{0,z} > 10 r_\mathrm{1}$. Following Refs.~\cite{Fuerst2018,Hirzler2020}, we propagate the particles using a fourth order step size adaptive Runge-Kutta method. The simulation ends when the separation between the Rydberg core and the ion is larger than $R_{0,z}$. We exclude events, in which the energy fluctuations between start and end of the collision $(E_\mathrm{init}-E_\mathrm{end})/E_\mathrm{init}$ exceeds $5\%$. Finally, we evaluate the binding energies of the two subsystems to extract the reaction channel and the quantum state. 

\begin{figure}[]
    \centering
    \includegraphics[width=\linewidth]{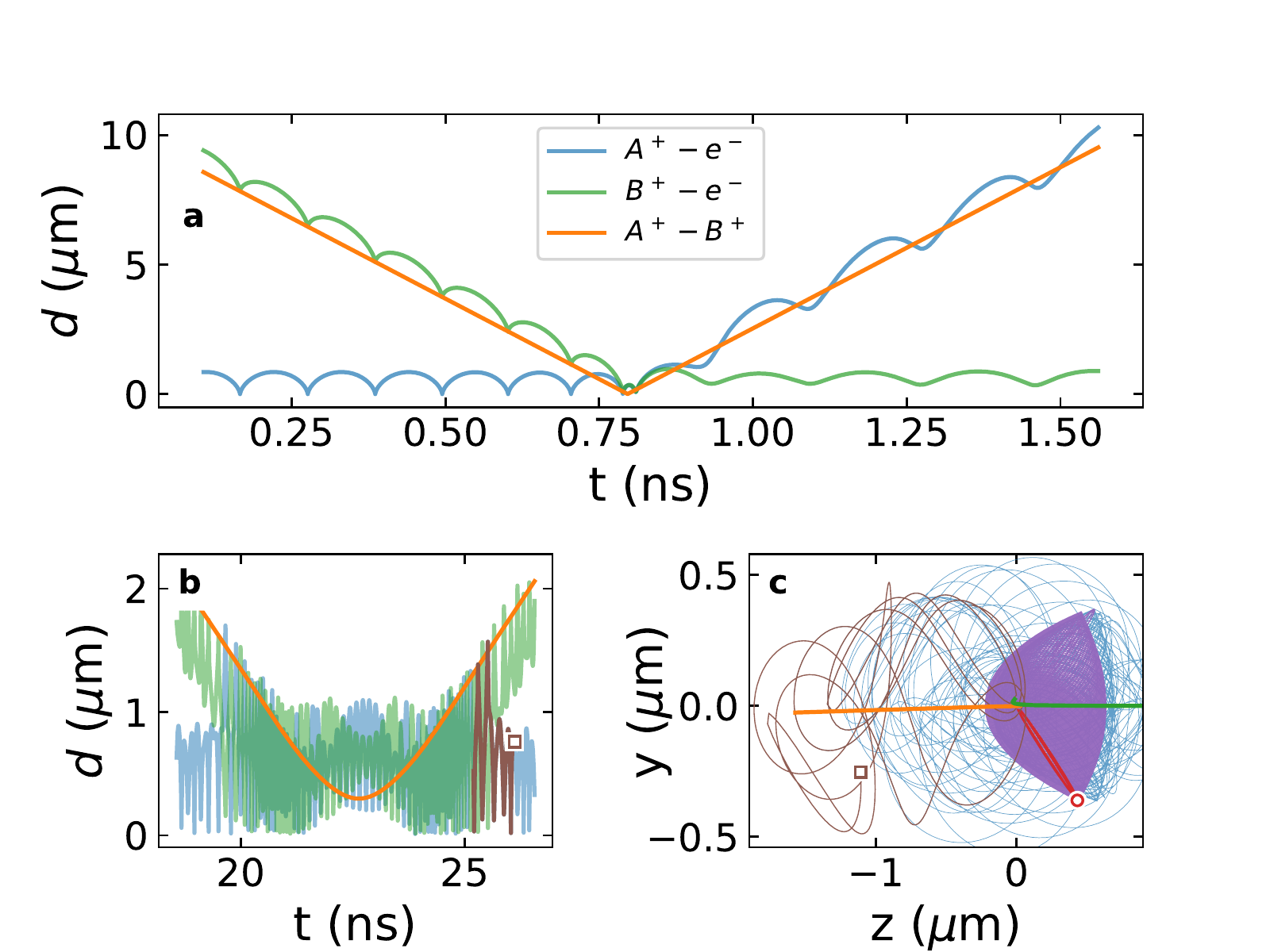}
        \caption{Trajectories for Li$^+$-Li$^*$ ($90D$) collisions from QCT simulations. Fast collision a) at $E_\mathrm{col}= 25000\,$K with (orange in the online version) Rydberg core-ion, (blue in the online version) Rydberg core-electron and (green in the online version) ion-electron distance $d$.
        Cold collision b) and c) at $E_\mathrm{col}=50\,$K. b) zoom in ($t\in[18,27]$ ns) with same coloring as a). c) Trajectories in z-y direction for $t \in [0, 26]\,$ns. Rydberg core (orange) initially at origin and ion (green in the online version) propagating in -z direction. Electron trajectory starting at (circle) t=0, (red in the online version) initial orbits, (purple in the online version) electron motion in far-field of the ion, (blue in the online version) continued propagation and (brown in the online version) final motion 
        before capture at $t\approx 26\,$ns.
        }
        \label{fig_trajectory}
\end{figure}

Trajectories for exemplary Rydberg-ion collisions are shown in Fig.~\ref{fig_trajectory}. a) Two particle distance $d$ between ($A^+-e^-$) Rydberg core-electron, ($B^+-e^-$)) ion-electron and ($A^+-B^+$) Rydberg core-ion for $E_\mathrm{col}=25000\,$K. During the collision, the electron is shared between the ionic cores for less than a period before it charge transfers. A much slower collision with $E_\mathrm{col}=50\,$K is shown in b), where the electron is shared for hundreds of orbits, before it ends up at one of the ions. It can be seen that in about $22.5\,$ns the distance of closest approach for the ionic cores is reached, as a result of the core repulsion. In c), the $z-y$ projection of the same trajectory is shown for ion, Rydberg core and Rydberg electron. The initial elliptical orbit is affected by the (near homogeneous) far-field of the ion.

Rydberg-ion collisions show three scattering channels:
\begin{itemize}
    \item[(i)] charge transfer $A^*(n,l)+B^+\rightarrow A^+ + B^*(n',l')$
    \item[(ii)] quenching ($n$-changing and $l$-mixing collisions in Rydberg physics) $A^*(n,l)+B^+\rightarrow A^*(n',l') + B^+$
    \item[(iii)] ionization $A^*(n,l)+B^+\rightarrow A^+ + B^+ + e^-$
\end{itemize}
We discriminate the reaction products by calculating the energy of the subsystems $A^+ + e^-$ and $B^+ + e^-$ after a collision. The probability for the scattering channel $\chi$ at impact parameter $b$ (opacity function) is obtained by Monte Carlo sampling of the initial conditions as
\begin{align}
	P_\chi(b)  &=  \frac{N_\chi(b)}{N(b)}\pm \delta_{N_\chi(b)}\\
	\delta_{N_\chi(b)} &= \sqrt{\frac{N_\chi(b) (N(b)-N_\chi(b))}{N(b)^3}},
	\label{eq_prob}
\end{align}
where $N$ is the total number of trajectories, $N_\chi(b)$ denotes the number of trajectories resulting in channel~$\chi$ and $\delta_{N_\chi(b)}$ is the error estimator. Subsequently, the reaction cross section is given by
\begin{equation}
\label{eq4}
    \sigma = 2\pi \int_0^{b_{\text{max}}} P_\chi(b)bdb,
\end{equation}
where $b_{\text{max}}$ is the maximum impact parameter at which the channel $\chi$ plays a role on the scattering. In this work, Eq.~(\ref{eq4}) is solved numerically.

\section{Results}

\begin{figure}[h!]
    \centering
    \includegraphics[width=\linewidth]{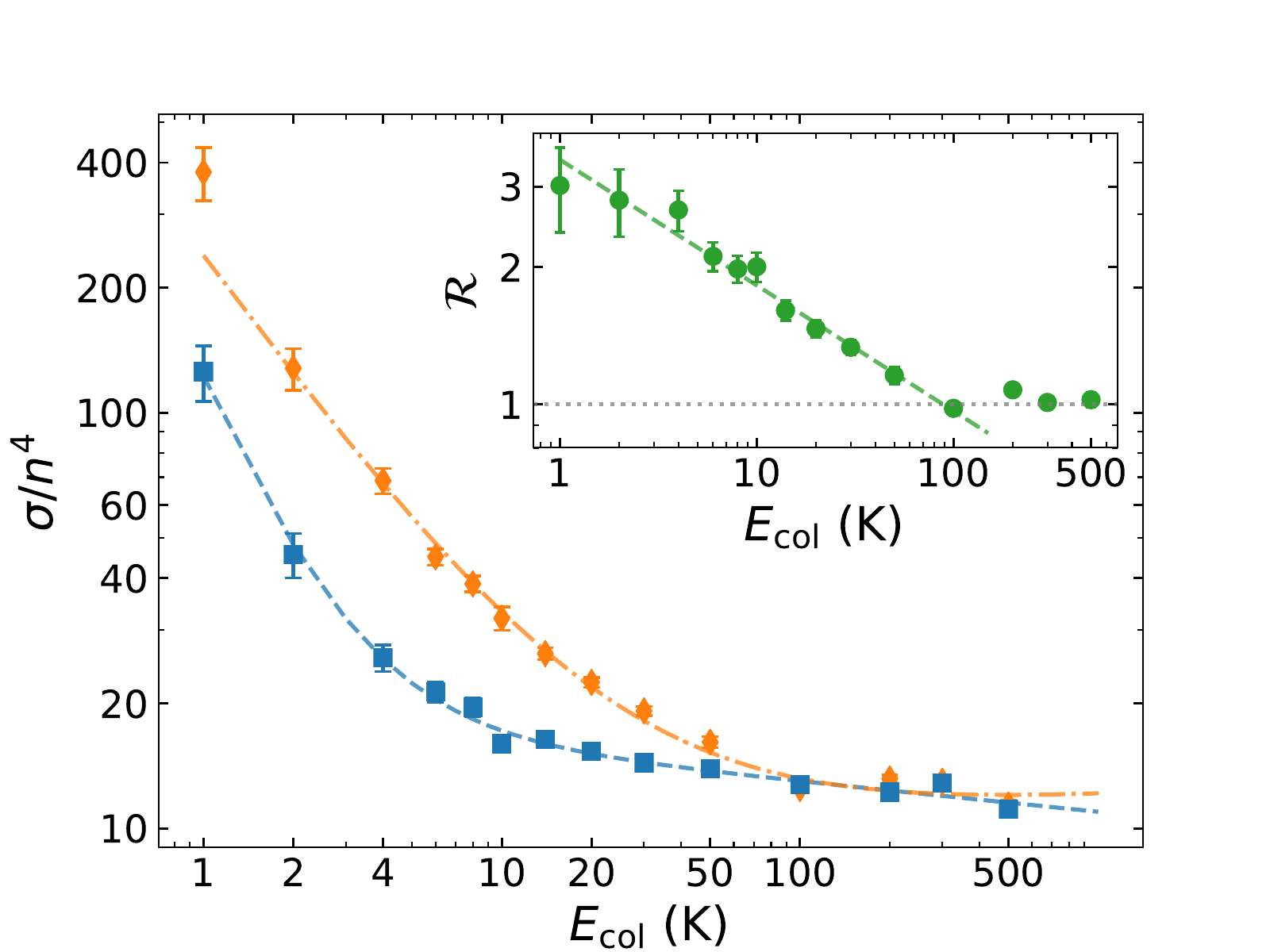}
        \caption{Charge exchange cross section for Li$^*_{n=90}$-Li$^+$ collisions. Numerical trajectory simulation (squares and blue in the online version) with Coulomb Rydberg core-ion interaction $\mathcal{A}=1.0$, and (diamonds and orange in the online version) without core-ion interaction $\mathcal{A}=0.0$. Dashed lines are fits with $\sigma/n^4=a(E_\mathrm{col})^b+c(E_\mathrm{col})^d$, with fit parameters $a,b,c$ and $d$. Inset: energy-dependent ratio $\mathcal{R}$ between the cross section neglecting core-ion interaction and the results including it. Dashed line (green online version) is a fit with $\mathcal{R}=a (E_\mathrm{col})^b$ for region $E_\mathrm{col}\leq 100\,$K, with free parameters $a$ and $b$. Errors originate from Eq.~\ref{eq_prob}.}
        \label{fig_cia}
\end{figure}

The cross section for charge exchange is investigated as a function of the collision energy using the QCT method outlined above. We start with the study of Rydberg-ion collisions in Li$^*$-Li$^+$ for $90D$ Rydberg states. In Fig.~\ref{fig_cia} the cross section for (squares) Coulomb core-repulsion ($\mathcal{A}=1$) and (diamonds) no core interaction are presented for energies reaching from $500\,$K down to $1\,$K~\footnote{Please note that at the lowest collision energies in this figure, the system is already in the Langevin regime. However, the current classical charge exchange processes may be helpful as a lower bound of the expected charge exchange cross section}. In this figure, we notice that the inclusion of Coulomb core-repulsion leads to smaller charge exchange cross sections for collision energies $\lesssim 100$~K. Indeed, this can be correlated with the results of the barrier-less model in which the inclusion of the ion-core repulsion leads to smaller charge exchange cross sections, as it has been argued in Section~\ref{CL}. In addition, we find that the charge exchange cross section shows the same trend as a function of the collision energy independently on the initial angular momentum state of the Rydberg electron.

To further examine this behavior, we have calculated the ratio $\mathcal{R}$ between the theoretical models, and the results are shown in the inset of Fig.~\ref{fig_cia}. Here, it is observed that both theoretical models agree with each other for collision energies $\gtrsim 100$~K. Therefore, 100~K seems to define a threshold energy from which the Rydberg-ion core repulsion does not play a role in the dynamics of charge exchange processes. Furthermore, we observe that the cross section strongly depends on the collision energy in stark contrast to the barrier-less models. In particular, the deviation from barrier-less models is more tangible for low-energy collisions.

In the case of Langevin collisions, following Eq.~(\ref{eq4}), one finds that $\sigma\propto \sqrt{\alpha}$. For the case at hand, taking into account that the polarizability of a Rydberg atom scales as $n^7$ it is easy to prove that $\sigma\propto n^{7/2}$. As a result the Langevin cross section scales less dramatically with $n$ than the present charge exchange model. In particular, assuming a 50$\%$ probability for charge exchange~\cite{Ostrovsky_1995}, the Langevin model results in larger cross sections e.g.  $\sigma_\mathrm{L}/n^4(500\,\mathrm{K})\gtrsim$ 100 and $\sigma_\mathrm{L}/n^4(100\,\mathrm{K}) \gtrsim 200$, which is expected due to the attractive force between the induced dipole and the ion.


\begin{figure}
    \center
    \includegraphics[width=\linewidth]{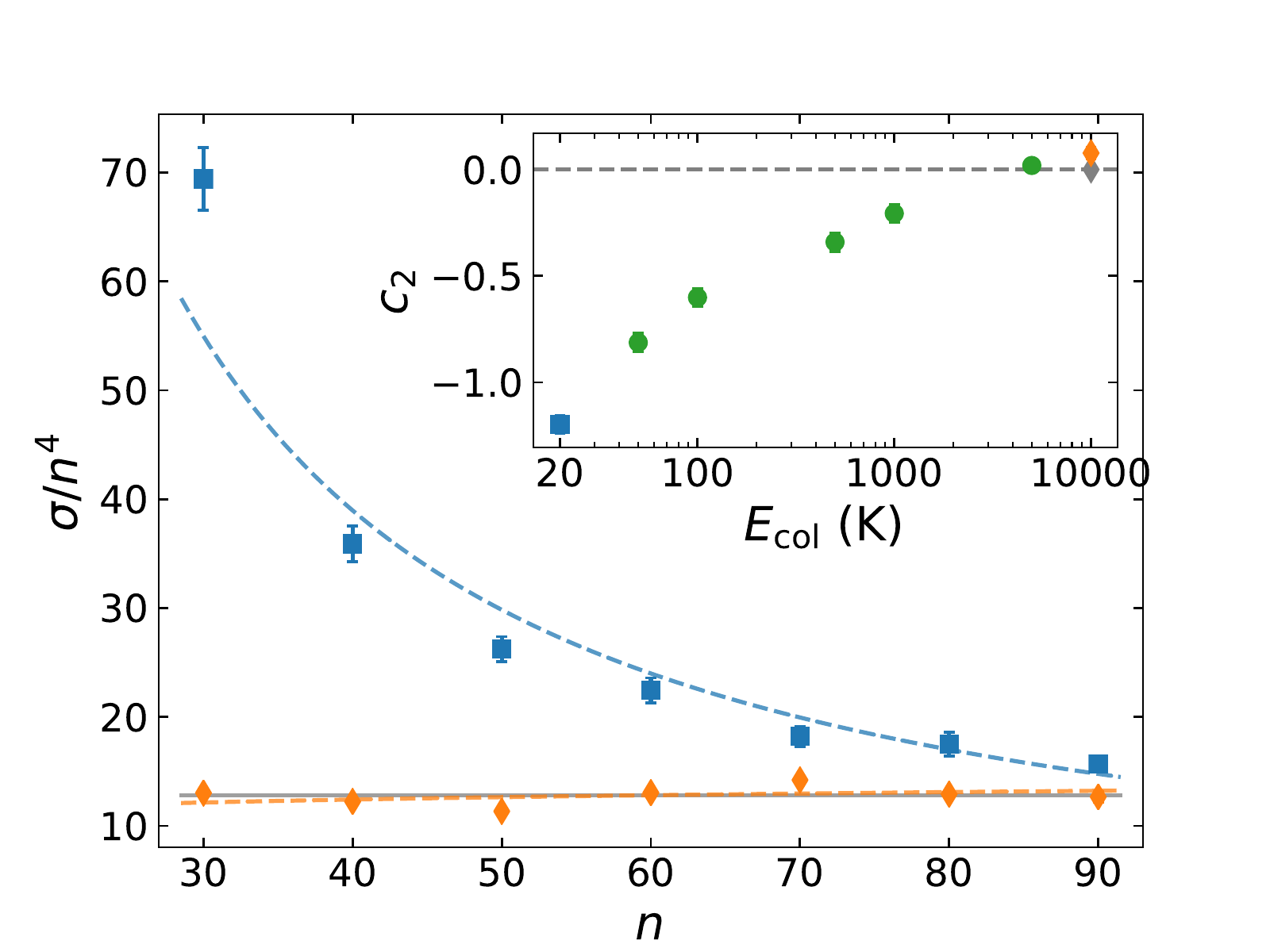}
    \caption{Charge exchange cross section for Li$^*$-Li$^+$ collision as a function of principal quantum number $n$ for (squares and blue in the online version) $E_\mathrm{col}=20\,$K and (diamonds and orange in the online version) $E_\mathrm{col}=10^4\,$K. Dashed lines are fits with Eq.~\ref{eq_fit_c2} and solid the line is a fit with $c_2=4$ fixed. The Inset shows the deviation of $c_2$ as a function of the collision energy, where $c_2=0$ corresponds to geometric scaling. Colors as in main plot. Errors originate from Eq.~\ref{eq_prob}.
    }
    \label{fig_n_principal}
\end{figure}

To elucidate the scaling of the charge exchange cross section with the principal quantum number $n$, we have computed the charge exchange cross section at a given collision energy as a function of $n$, and the results are presented in Fig.~\ref{fig_n_principal}. For (diamonds) high-energy collisions with $E_\mathrm{col}=10000\,$K we find $\sigma \propto n^4$ corresponding to the geometric cross section given by the barrier-less model (see Section~\ref{CL}), whereas for (squares) low-energy collision at $E_\mathrm{col}=20\,$K a deviation from the geometric scaling law is visible. Therefore, the geometric cross section is only a good estimation of the charge exchange cross section for high-energy collisions. To quantify this observation, we repeat the simulation for various collision energies and the resulting charge exchange cross section is fitted as
\begin{equation}
    \sigma/n^4=c_1n^{c_2},
    \label{eq_fit_c2}
\end{equation}
where $c_i$ are fitting parameters. The resulting parameter $c_2$ is shown in the inset of Fig.~\ref{fig_n_principal} as a function of the collision energy. As a result, the geometric scaling is recovered for energies $\gtrsim 5000\,$K where $c_2\approx0$. Thus, this collision energy determines the lowest energy in which the barrier-less models are reliable estimators of the charge-exchange cross section. For low-energy collisions, the $n$-dependence of the scaling parameter $c_2$ changes significantly with the collision energy. In particular, the lower the collision energy, the less steep is the cross section's dependence with $n$. Indeed, for a collision energy of 20K the cross section scales  $\propto n^{3}$. 

\begin{figure}
    \center
    \includegraphics[width=\linewidth]{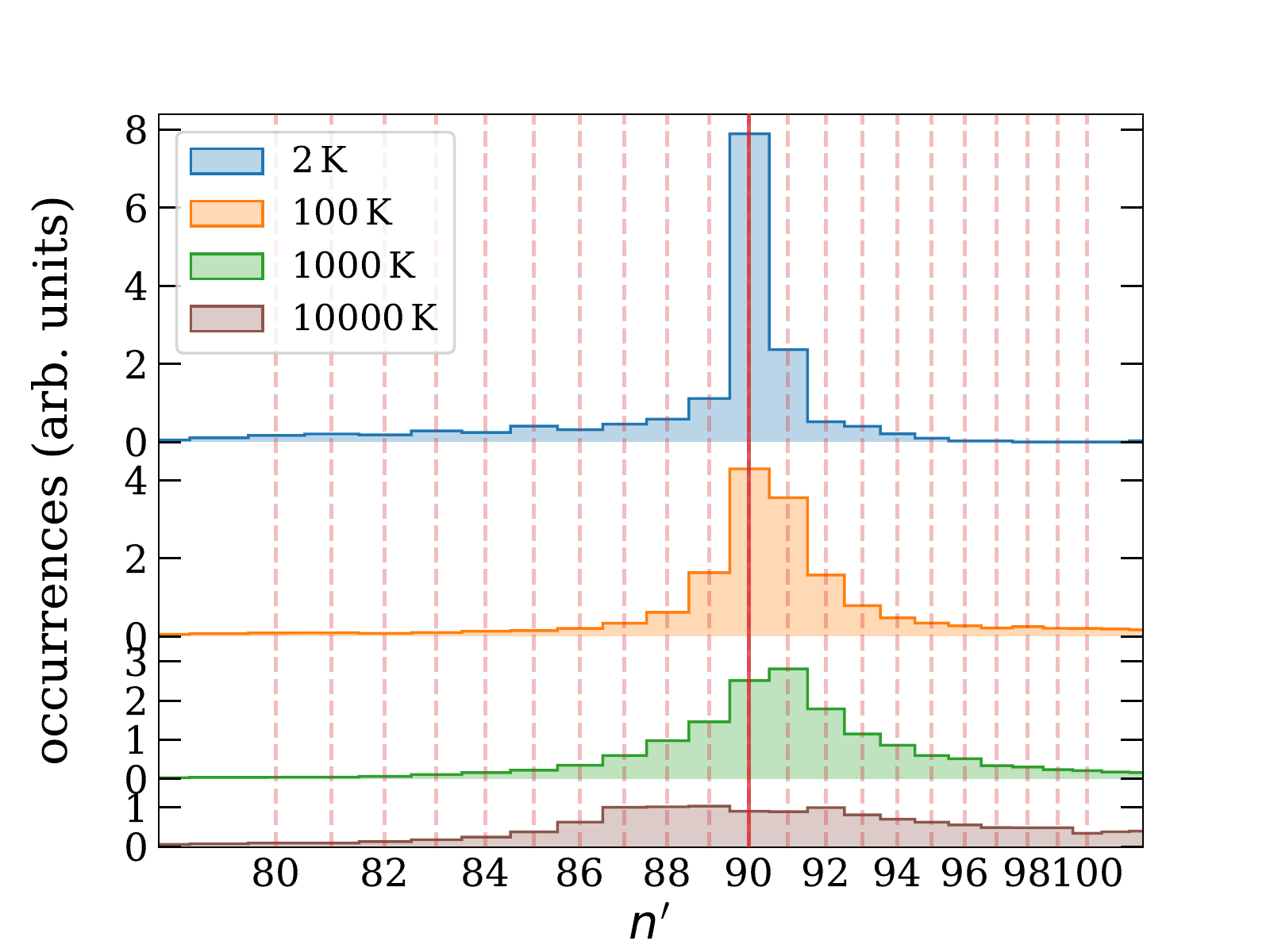}
    \caption{Principal quantum number $n'$ distribution for charge exchange collisions in Li$^*$-Li$^+$, initially in $n=90$ and collision energies $E_\mathrm{col}= 2-10000\,$K as indicated in the legend.}
    \label{fig_n_distribution}
\end{figure}

Next, we study the final state distribution of the Rydberg atom after charge transfer. To this end, focusing on Li($nD$)-Li$^+$ collisions, we compute the final state distribution of principal quantum numbers $n'$ of the product Rydberg atoms as a function of the collision energy, and the results are shown in Fig.~\ref{fig_n_distribution}. The figure shows that the populated states after charge transfer are closely distributed around the Rydberg atom's initial state, i.e., $n=90$, independently of the collision energy. However, the distribution broadens for higher collision energies. For instance, for a collision energy of $10^4\,$K (bottom distribution), the distribution is widely spread about $n\pm4$ (FWHM) around $n'=90$. Furthermore, the product state distribution shows a clear asymmetry towards higher principal quantum numbers, which can be rationalized by considering that the energy difference between consecutive Rydberg states scales as $n^{-3}$. Therefore, higher principal quantum numbers are closer in energies than lower ones, which leads to higher population and spread. 

\begin{figure}
    \center
    \includegraphics[width=\linewidth]{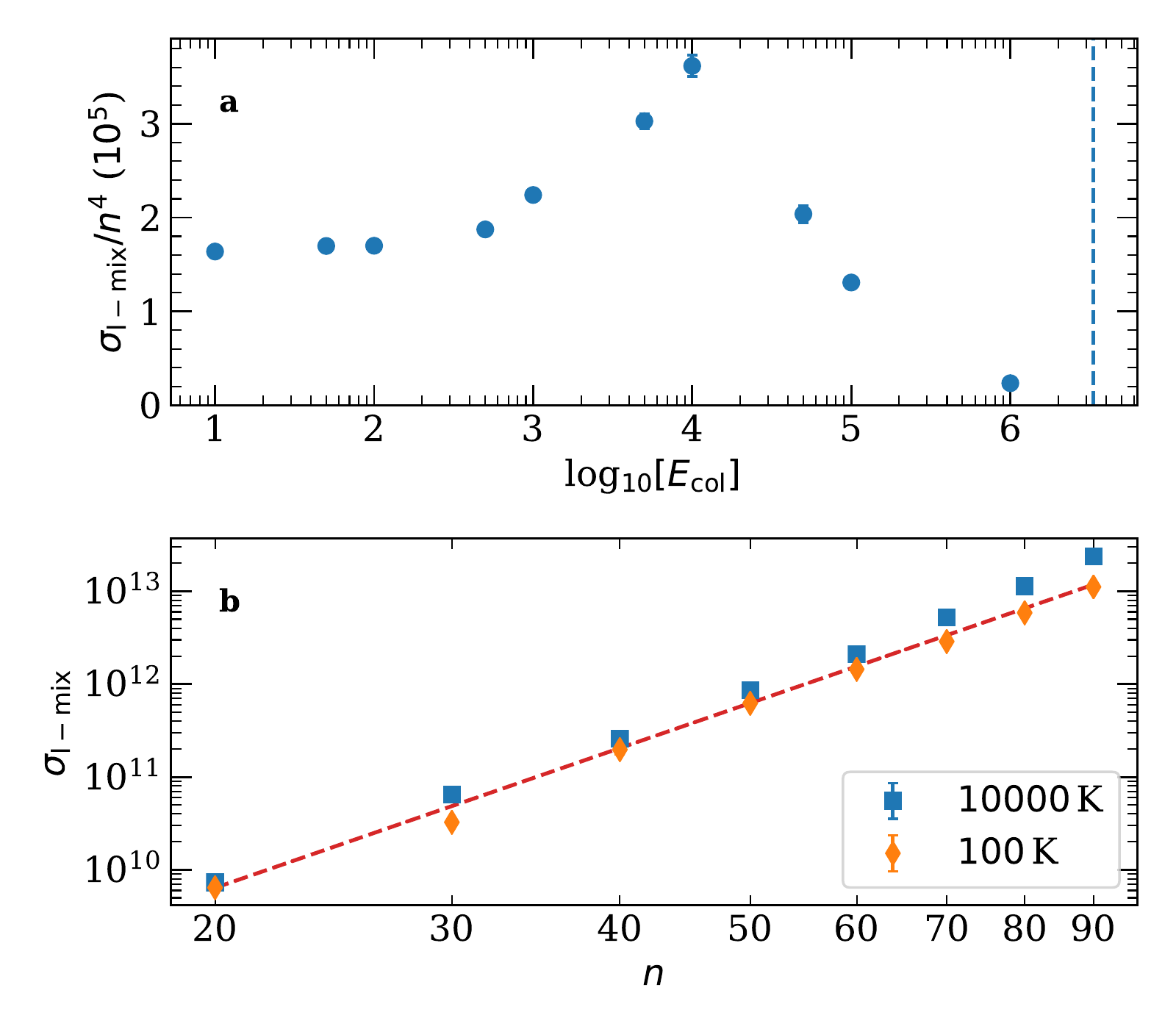}
    \caption{Angular momentum number $l$-mixing cross section as a function of (a) collision energy. The dashed (blue in the online version) line indicates the characteristic electron velocity $v_e = 1/n$. Effect of the initial principal quantum number on the $l$-mixing cross section. Dashed (red in the online version) line shows a power law $\sigma_\mathrm{l-mix}\propto n^5$.}
    \label{fig_l-mixing}
\end{figure}

Ion-Rydberg collisions may lead to quenching collisions as summarized in Section~\ref{Sec_theory}. These comprise all collisions, after which the electron is confined to its original core and $n$, $l$ both or only one of them has changed. We simulate the quenching cross section for various collision energies and present the results in panel a) of Fig.~\ref{fig_l-mixing} with the Rydberg atom initially in $90D$. It is worth pointing out that for these calculations, the initial Rydberg-ion distance was increased by an order of magnitude in comparison with charge exchange ones to account for convergence of the final $l$-state. For $E_\mathrm{col}= 10\,$K we find $\sigma/n^4\approx1.6(0.6)\times 10^5$, exceeding the charge exchange cross section by about four orders of magnitude. This giant cross section results from the tiny energy gap between inelastic channels and the ion's electric field's effect on the Rydberg electron orbit. For hot collisions with energies approaching the characteristic velocity of the Rydberg electron $v_e$ the cross section decreases in agreement with results from hot Rydberg-ion experiments~\cite{GallagherBook}.

We vary the initial $n$-state of the Rydberg atom for a fixed collision energy and simulate the quenching cross section as it is presented in Fig.~\ref{fig_l-mixing} b). For $100\,$ K (diamonds), the cross section increases with $n$, following a power-law with a scaling exponent of about five, shown as a dashed line. Our results are in agreement with hotter collisions with projectile velocities similar to the electron orbit velocity found in Na$^+$-Na$^*$ collisions~\cite{MacAdam_1981}.


\begin{figure}
    \center
    \includegraphics[width=\linewidth]{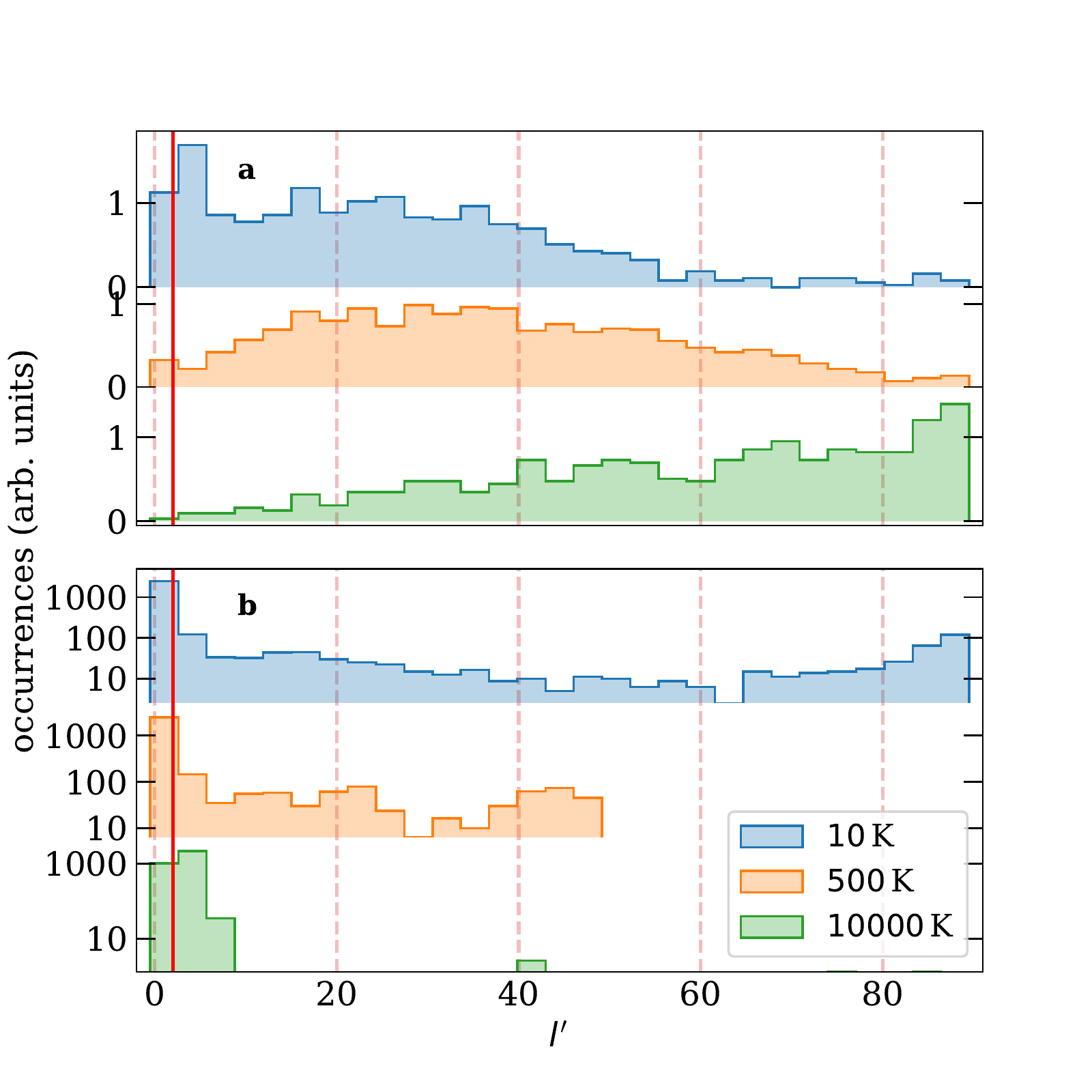}
    \caption{Angular quantum number $l'$ distribution for collisions with $n'=n$ either a) from charge exchange or b) with the electron remaining bound to its original core. Initial state is $90D$ and with collision energies $E_\mathrm{col}= 10-10000\,$K as indicated in the legend.}
    \label{fig_l-mixing_hist}
\end{figure}

Next, we study the distribution of the final angular momentum quantum number, $l'$, for collisions with $n'=n$ ($l$-mixing collisions) for different collision energies as it is shown in Fig.~\ref{fig_l-mixing_hist}. For charge exchange processes, panel a), we find for $10\,$K (top histogram) a widespread distribution of angular momenta below $l\approx60$, whereas higher $l'$ are sparsely found. For hotter collisions at $500\,$K (medium histogram) the $l'$-distribution migrates to higher values, and for $1000\,$K (bottom histogram) we find the peak of the population distribution at the highest allowed states. This is in correspondence with the localization of $n'$ for low-energy collisions, as shown in Fig.~\ref{fig_n_principal}. On the contrary, as shown in panel b), $l$-mixing collisions show a narrower distribution of angular momentum states at higher collision energies, in correspondence with dipole's dominant role allowed transitions. Whereas, for lower collision energies, the dipole selection rule does not dominate the dynamics, leading to a wider distribution of the final angular momentum states~\cite{GallagherBook}.

\begin{figure}
    \center
    \includegraphics[width=\linewidth]{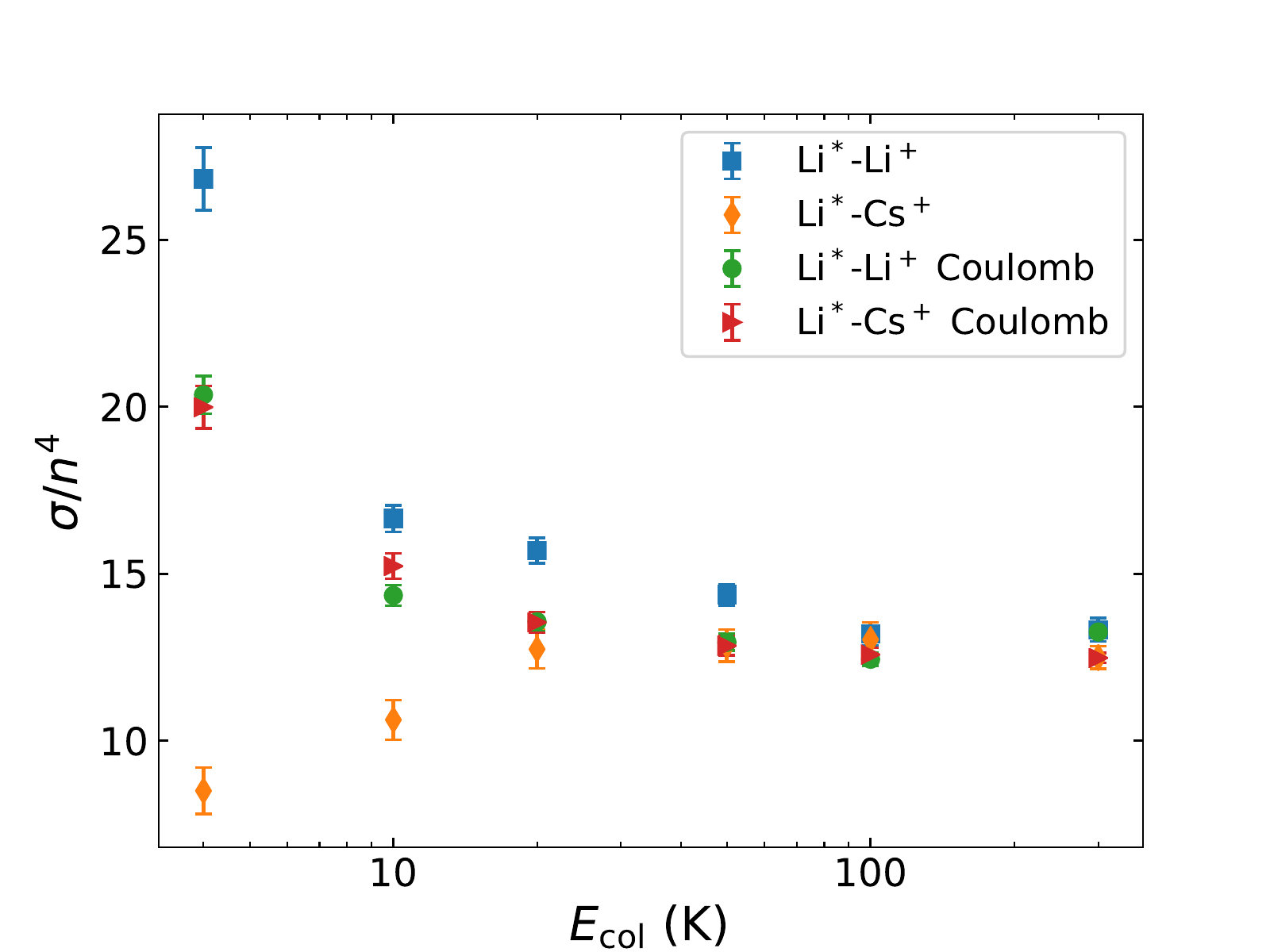}
    \caption{Charge exchange cross section for Li$^*$-Li$^+$ and Li$^*$-Cs$^+$ ($90D$) collisions using (green and red, respectively in the online version) Coulomb or (blue and orange, respectively in the online version) pseudo potential to model the electron-ion interaction. Errors originate from Eq.~\ref{eq_prob}}.
    \label{fig_mass_ratio}
\end{figure}

So far we have assumed that the interaction between the Rydberg electron and the ions (core of the Rydberg atom and ion) is dominated by a pseudo potential including effects of core valence electrons. However, the same scenario can be studied assuming a pure Coulomb interaction~\cite{Ostrovsky_1995,Becker_1984,Olson}. To elucidate the role of the pseudo potential we simulate low-energy Li$^*$-Li$^+$ collisions assuming a pure Coulomb interaction and a pseudo potential and present the obtained charge exchange cross section as a function of the collision energy in Fig.~\ref{fig_mass_ratio}. In this figure, it is noticed that pure Coulomb interaction (circles) leads to lower charge exchange cross sections than when a pseudo potential (squares) is employed. This discrepancy can be exploited to elucidate the effects of the pseudo potential directly on scattering observables. In particular, studying Rydberg-ion charge exchange process at low-energy collisions would be possible to identify and possibly isolate the effects of the pseudo potential by comparing the measured charge exchange cross section with the classical calculations presented in this work. 


Finally, we study non-resonant charge exchange processes. In particular,  we simulate Li$^*$-Cs$^+$ collisions for Coulomb potential and for the pseudo potential for which we adjust the parameters of Eq.~\ref{eq:param_model} for the electron-Cs$^+$ interaction. The obtained cross sections are shown in Fig.~\ref{fig_mass_ratio} for (triangles) Coulomb and (diamonds) pseudo potential. The first observable effect is that the non-resonant charge exchange cross section is systematically lower than the resonant one, as we expected. In particular, the difference is more palpable at low-energy collisions $E_\mathrm{col}\lesssim 10\,$K. This effect can be understood by considering that in the case of the pseudo potential, the interaction of e$^-$-Li$^+$ and e$^-$-Cs$^+$ is different, prohibiting resonant charge exchange. However, if one assumes a pure Coulomb interaction, the resonant and non-resonant charge exchange cross section agree with each other since the electron-ion interaction is the same, and only the reduced mass varies. The second observation is to realize that for collision energies $\gtrsim 100$~K, the difference between resonant and non-resonant charge-exchange vanishes in agreement with the results of Fig.~\ref{fig_cia}.

\section{Summary and conclusions}
A study on cold Rydberg-ion collisions has been presented. We have identified the parameter space for the quantum regime, Langevin regime, and classical exchange regime, which reveals classical models' applicability for the coldest energies in Rydberg-ion collisions. For cold collisions, our numerical results show a strong effect of the core repulsion on the charge exchange cross section $\lesssim 100\,$K. At similar temperatures, the electron-ion interaction model becomes essential, and we see differences in the charge exchange cross section using a pseudo potential or a Coulomb potential. Furthermore, we have found a deviation from the geometric scaling law predicted by the barrier-less model for collisions at energies below $\lesssim 5000\,$ K. In addition, we have computed $l$-mixing cross sections in an energetic regime never explored before. As a result, we have been able to extend previous estimations on the cross section's dependence with the principal quantum number and the distribution of the final angular momentum states of the Rydberg.

Our results indicate that it is possible to fit pseudo potentials by comparing experimental resonant charge exchange cross section measurements with the classical model presented here. In particular, it will be interesting to study cold Rydberg-ion collisions experimentally with respect to understanding the cores' interaction potential. Especially, ions in higher groups of the periodic table could deepen our description of atomic potentials beyond alkalies. A particular interest may be found in earth alkali ions, abundant in  cold atom-ion hybrid experiments. While atomic potentials can be conveniently obtained using polarizability measurements, low-energy Rydberg atom-ion collisions might be a supportive approach to access atomic potentials.

\section{Acknowledgemens}

We thank Dr. Rene Gerritsma for helping in the early stages of this project and his support and comments on our work. We like to acknowledge Thomas Secker and Dr. Henning F\"urst for fruitful discussions. This work has been supported by the Netherlands Organization for Scientific Research (
Start-up grant 740.018.008 (H.H.), and Vrije Programma 680.92.18.05) (H.H. and J.P.-R.).




%

\end{document}